\documentclass{ws-procs975x65}

\newcommand{\beq}{\begin{equation}}
\newcommand{\eeq}{\end{equation}}
\begin{document}

\title{Networks as Renormalized
 Models for Emergent Behavior in Physical Systems}

\author{MAYA PACZUSKI}

\address{Perimeter Institute for Theoretical Physics, Waterloo, Canada, N2L 2Y5 \\ and\\
         Department of Mathematics,  Imperial College London, London, UK  SW7 2AZ \\
         E-mail: maya@ic.ac.uk}  

\maketitle

\abstracts{Networks are paradigms for describing complex biological,
social and technological systems.  Here I argue that networks provide
a coherent framework to construct coarse-grained
models for many different physical systems. To elucidate these ideas,
 I discuss two long-standing
problems. The first concerns the structure and dynamics of 
magnetic fields in the solar corona, as exemplified by sunspots that
startled Galileo almost 400 years ago. We discovered that the magnetic
structure of the corona embodies a scale free network, with spots at
all scales.  A network model representing the three-dimensional
 geometry of magnetic fields,
 where links rewire and nodes merge when
they collide in space, gives quantitative agreement with available
 data, and suggests new measurements.  Seismicity is
addressed in terms of relations between events without imposing space-time
windows. A metric estimates  the correlation between any
two earthquakes.  Linking strongly correlated pairs, and ignoring
pairs with weak correlation organizes the spatio-temporal process into
a sparse, directed, weighted network.
 New scaling laws for seismicity are found.
  For instance,  the aftershock decay rate decreases as
$\sim 1/t$ in time up to a correlation time, $t_{\rm omori}$.
An estimate from the data gives $t_{\rm omori}$
  to be  about one year for small magnitude 3 earthquakes, about 1400
years for the Landers event, and roughly 26,000 years for the
earthquake causing the 2004 Asian tsunami. Our results confirm Kagan's conjecture
that aftershocks can rumble on for centuries.}

\section{Introduction}

 A fundamental problem in physics, which is not always recognized as
being ``a fundamental physics problem'', is how to mathematically
describe emergent phenomena.  It seems hopeless for many reasons to
make a theory of emergence that harmonizes all
scales, from the Planck scale  to the size of our
Universe, and includes life on Earth with its manifest details, such as
bacteria,  society, or ourselves as individual personalities.
That would be a true theory of everything (TToE). (For a discussion
see Refs.~[1,2].)

  However, a reasonable aim is to describe how entities or excitations
with their own effective dynamics develop from symmetries, conservation laws
 and
nonlinear interactions between elements at a lower level.  Some famous
examples in statistical physics are critical point fluctuations,
avalanches in sandpiles,\cite{btw} vorticity in turbulence,\cite{turbulence} or the distribution
of luminous matter in the Universe.\cite{pietronero,bak_chen,bak_pac}  Contemporary work in quantum
gravity suggests that both general relativity and quantum mechanics
 may emerge from coarse graining a
low energy approximation to the fundamental causal histories.  These
histories are sequences of changes in graphs, that may be nonlocal or
small-world networks.\cite{fotini}  
Similar sets of questions crop up across the
board.  How do you get qualitatively new structures and dynamics from underlying
laws?

An important distinction appears between equilibrium and far from
equilibrium systems.  Roughly speaking, most equilibrium systems are
complex in the same way. They exhibit emergent behavior at critical
points with fluctuations governed by symmetry
principles, etc. Non-equilibrium systems, however, seem complex in a
myriad, different ways. 

However, a variety of indicators point to principles of organization
for emergent phenomena far from equilibrium. 
Various types of scaling behaviors in physical systems
(scale invariance,\cite{bassler}  scale covariance,\cite{dubrulle,chen_bak} etc.) can be quantitatively predicted
 using coarse-grained models. After all, the underlying 
equations typically govern at length and time scales well below those
where observations are made.    The key is to
capture the dynamics of larger scale entities, or "coherent
structures",\cite{chang} 
and use those as building blocks to model the whole
system. Ideally, renormalized models may be derived
from the underlying equations, but it is not clear that this is always possible.
  Even without an explicit derivation, though,
once such a model is born out
in a specific system, by subjecting it to falsifiable tests, it may
also connect to other physical situations with similar, or even
different underlying laws.

Nowadays, computational science
tends to emphasize studies of bigger and bigger systems
with more and more details. That is unlikely, by itself, to lead to
any better understanding of emergence, and also can easily be
demonstrated to be fruitless for many interesting problems in physics,
like those discussed here.  There are simply too many degrees of
freedom coupled over too long times, compared to the microscopic time.
That doesn't mean that these problems are unsolvable through
computational methods though. We  must use a different starting
point. 

Complex networks have been intensively investigated recently as descriptions
of biological, social and technological phenomena.\cite{albert,newman}
In fact, a sparse network expresses
coarse-graining in a natural way, since the few links present
highlight relevant interactions between effective degrees of freedom,
 with all other nodes and links deleted.  Then
renormalization may proceed further on the network alone by grouping tightly
coupled nodes or modules together and finding the
interactions  between those new effective degrees of freedom.
Understanding processes of
network organization, perhaps through an information theory of complex
networks,\cite{adele} is (arguably) necessary to make progress toward theories of emergence
in physical systems.

In order to demonstrate the wide applicability of these ideas in diverse
contexts and at different levels in our ability to describe physical phenomena, I present two  distinctive examples of networks as empirical
descriptions for physical systems.  
First, I discuss the coronal magnetic field and show
 that much of the important physics can be captured with a network
 where nodes and links interact in space and time with local
 rules. In this case, the network is an abstraction of the geometry of the
 magnetic fields.   We use insights gained from
 studying the underlying equations, and a host of observations from
 the Sun to determine a minimal model.\cite{modelprl,Hughes:longpaper,heavenly}
 
 Second, I discuss a new approach to  seismicity based solely 
 on
 relations between events, and not on any fixed space, time or magnitude
 scales.
  Earthquakes are represented
 as nodes in the network, and strongly correlated pairs are linked. A  sparse,
 directed network of disconnected, highly clustered graphs emerges. The ensemble
 of variables and their relations on this network reveal new scaling laws
 for seismicity.

Our network model of coronal magnetic fields is minimal in that
 if any of its five basic ingredients  are deleted then
 its behavior changes and fails to agree with observations.
  However, its rules can be changed in many ways, for instance by
 altering parameters, or adding interactions without modifying most statistical
 properties.   Although the model is not explicitly constructed
 according to a formalism based on symmetry principles, relevant
 operators, and general arguments used for statistical field theories,
 it appears to have comparable robustness and fixed point properties.
 Lastly, the model is falsifiable. We have made numerous predictions
 for observables, as well as suggesting new quantities to be measured.
 In fact, studying its behavior led us to re-analyze previously
 published coronal magnetic field data, and reveal the scale-free
 nature of magnetic ``concentrations'' on the visible surface of the
 Sun.

\section{The Coronal Magnetic Field}

The Sun is a magnetic star.\cite{zirker}  Like Earth, matter density at its surface
drops abruptly, and a tenuous outer atmosphere, the corona, looms
above.  The surface, or photosphere, is much cooler than both the
interior of the Sun, and the corona. For this reason, only 
magnetic fields at or near the surface have been directly
measured. Several mechanisms have been proposed for coronal heating
including nanoflares.\cite{parker_nano} Like bigger, ordinary flares, these may be caused
by sudden releases of magnetic energy from reconnection.  Reconnection
occurs when magnetic field lines rapidly rearrange
themselves. Fast
reconnection is a threshold process that occurs when magnetic field
gradients become sufficiently steep.\cite{parker_nano,parker,lh}

In the convective zone below the photosphere, temperature gradients
drive  instabilities. Moving charges in the plasma create
magnetic fields.  Rising up, these fields pierce the photosphere and
loop out into the corona.  The pattern of flux on the photosphere and
in the corona is not uniform, though. Flux is tightly bundled
into long-lived flux tubes that attach to the photosphere at
footpoints.  These flux loops survive for hours or more, while the lifetimes
of the granules on the photosphere is minutes. 

\begin{figure}
\centering
\includegraphics[width=350pt]{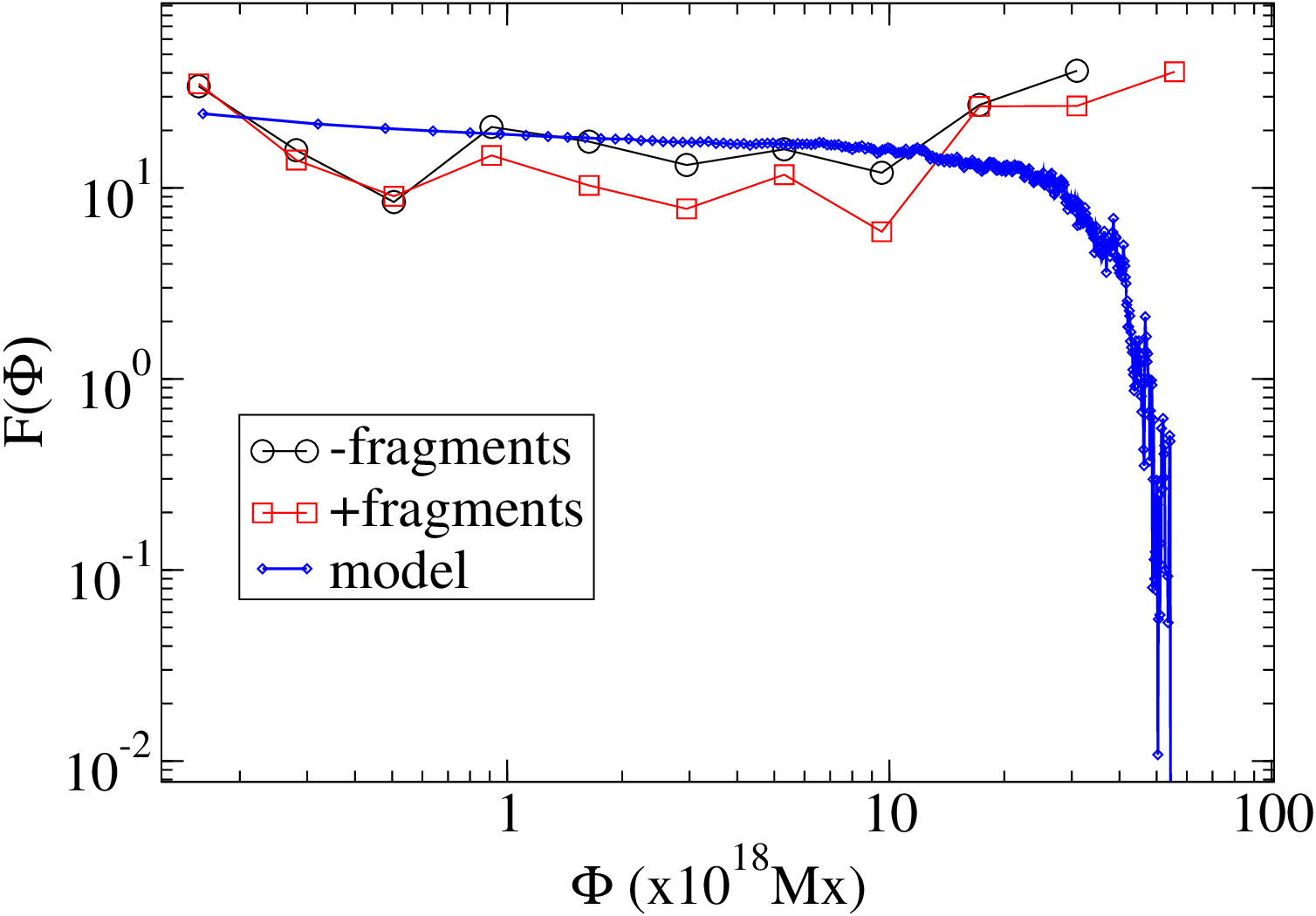}
\caption{Results demonstrating the scale free network of coronal
magnetic flux and comparison with results from numerical simulations
of our self-organizing network.  For the concentration data
$F(\Phi)=constant\times P(\Phi)(\Delta \Phi) \times (\Phi)^{1.7}$,
where $P(\Phi)(\Delta \Phi)$ is the normalized number of magnetic
concentrations in bins of size $\Delta \Phi = 1.55\times 10^{17}$ Mx,
obtained by re-analyzing the measurement data originally shown in
Figure 5 of Ref. [24].
  The model data shown represents the
probability distribution, $P(k_{foot})$, for number of loops, $k_{foot}$, connected to a
footpoint. This has been rescaled so that one loop, $k_{foot}=1$,
equals the minimum threshold of flux, $1.55 \times 10^{17}$ Mx. The
cutoff at large $\Phi$ in the model data is a finite size effect that
can be shifted to larger values or smaller ones by changing the size
of the system.}
\end{figure}

Footpoints aggregate into magnetic ``concentrations'' on the
photosphere.  Measuring these concentrations provides a quantitative
picture that can be compared with theory.  The strongest, and
physically largest concentrations are sunspots, which may contain more
than $10^{22}$~Mx.\cite{maxwell} The intense magnetic fields in these
regions cool the plasma, so they appear dark in the visible
spectrum. The smallest resolvable concentrations above the current
resolution scale of $\approx 10^{16}$~Mx are ``fragments''. 
Solar physicists have constructed elaborate theories where at
each scale a unique physical process is responsible for the dynamics
and generation of magnetic concentrations, e.g. a ``large scale
dynamo'' versus a ``surface dynamo'' etc. These theories predict an exponential
distribution for concentration sizes.

\subsection{Coronal Fields 
Form a Scale Free Network}

David Hughes and I re-analyzed\cite{Hughes:longpaper,heavenly}
  previously published
data sets reporting the distribution of concentration sizes.\cite{close}   As shown in Fig.~1, we discovered that
 the distribution is
scale free over the entire  range of
 measurement.  The probability to
have a concentration with flux $\Phi$,
$P(\Phi) \sim \Phi^{-\gamma}$ with $ \gamma \approx 1.7$, 
as indicated by the
flat behavior of $F(\phi)$ in Fig.~1.  Similar results were found using other
data sets.\cite{Hughes:longpaper}

\subsection{The Model}

Results from numerical simulations of our network model are also shown
in Fig.~1. The only calibration used
 (which is unavoidable) was to set the minimal unit of flux
in the model equal to the flux threshold of the measurement. 
 How did we get such good agreement without solving
any of the plasma physics equations?

Considering the long-lived flux tubes as the important coherent
structures, we treated the coronal magnetic field as made up of
discrete interacting loops embedded in three dimensional space.\cite{modelprl}  Each
 directed loop traces the mid-line of a flux tube,
and is anchored to a flat surface at two opposite polarity footpoints.
A footpoint locates the center of a magnetic concentration, and is
considered to be a point.  A collection of these loops and their
footpoints gives a distilled representation of the coronal magnetic
field structure. Our network model is able to describe the three dimensional
geometry of fields that are
very complicated or interwoven.  

The essential ingredients, which must be included to agree with
observations are: injection of small loops,
 submergence of very small loops, footpoint
diffusion, aggregation of footpoints, and reconnection of loops. Observations
indicate that all of these physical
processes occur in the corona.\cite{parker_nano,parker,close} 
Loops injected at small length scales are stretched and shrunk as
their footpoints diffuse over the surface. Nearby footpoints of the
same polarity aggregate, to form magnetic fragments, which can
themselves aggregate to form ever larger concentrations of flux. Each
loop carries a single unit of flux, and the magnetic field strength at
a footpoint is given by the number of loops attached to it. The number
of loops that share a given pair of footpoints measures the strength
of the link. The link strengths also have a scale-free distribution
with a steeper power law than the degree distribution of the
 nodes, or concentrations.  Also, the
number of nodes that a given node is connected to by at least one loop
is scale-free. Both of these additional claims could also be tested against
observations.\cite{Hughes:longpaper,heavenly}

Loops can reconnect when they collide, or cross at a point
 in three dimensional space above the surface. The
flux emerging from the positive footpoint of one of the reconnecting
loops is then no longer constrained to end up at the other footpoint
of the same loop, but may instead go to the negative footpoint of the
other loop. This occurs if the rewiring lowers the combined loop length.
The loops between the newly paired footpoints then both relax  to a
semi-circular shape.  Reconnection allows footpoints to exchange
partners and reshapes the network, but it maintains the degree  of each
footpoint.

\begin{figure}
\centering
\includegraphics[width=350pt]{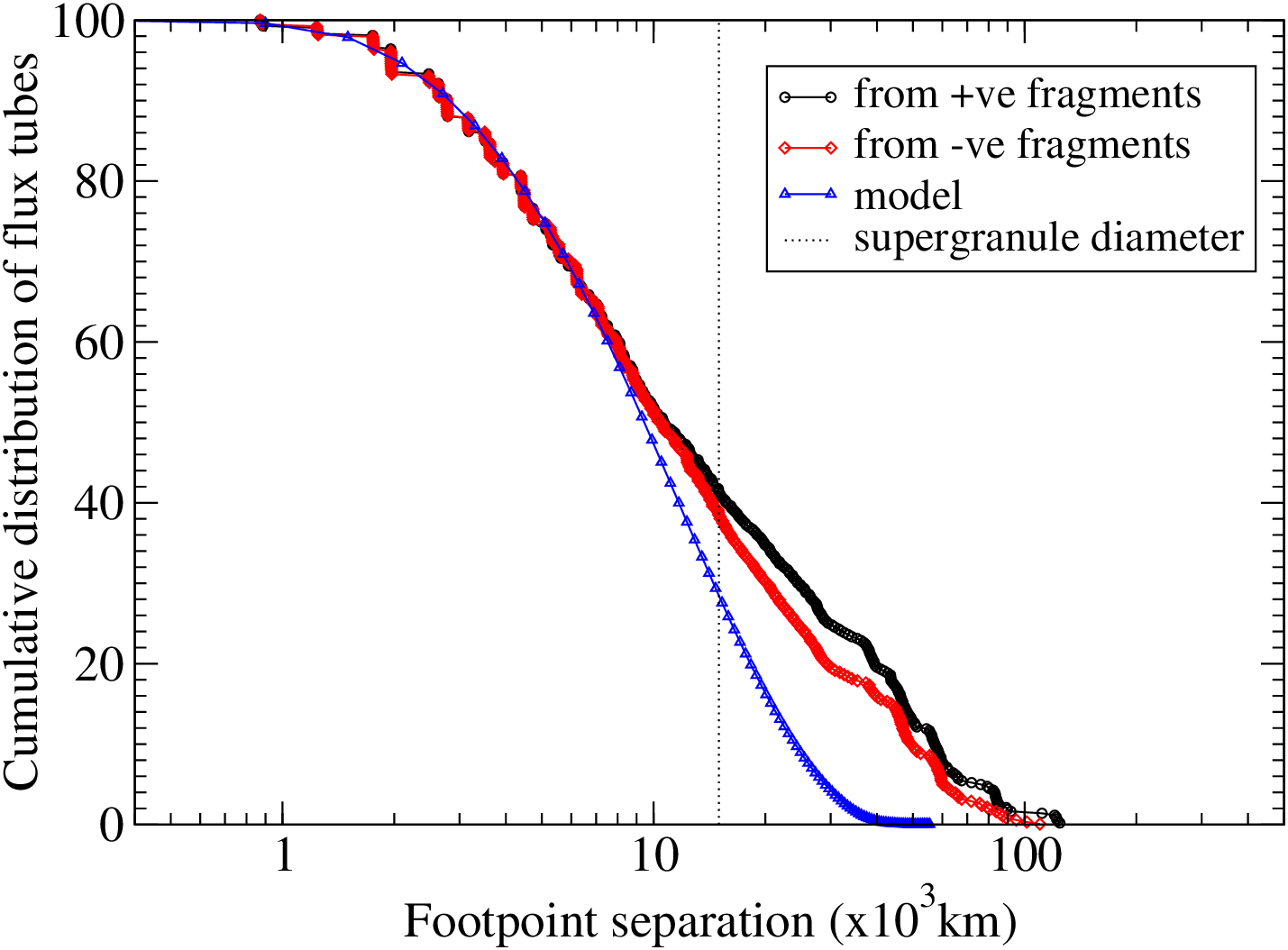}
\caption{The cumulative percentage of footpoint pairs 
separated by a distance on the photosphere
 larger than $d$. The flux tube data
corresponds to Figure 6c in Ref. [24].
 The model data has been scaled such
that one unit of length is equal to 0.5Mm.}
\end{figure}

If rewiring occurs,  one or both loops may need to cross
another loop. A single reconnection between a pair of loops
can trigger an avalanche of  reconnection.
 Reconnections occur instantaneously compared to the diffusion
of footpoints and injection of loops. It may also happen that due to
reconnection or footpoint diffusion, very small loops are
created. These are removed from the system.  Thus the collection of
loops is an open system, driven by loop injection with an outflow of
very small loops.  From any initial condition, the system
 reaches a steady state where the loops self organize into
a scale free network. As shown in Fig.~1 , the number of loops,
$k_{foot}$, connected to any footpoint is distributed as a power law
$$ P(k_{foot})\sim k_{foot}^{-\gamma} \, \, {\rm \ with \ \ }
 \gamma = 1.75 \pm -0.1 \quad .$$

\subsubsection{Further predictions of the network model}

The distribution of distances, $d$, between footpoint pairs attached to the
same loop can also be calculated and compared with measurement data as
shown in Fig.~2. Indeed, by setting one unit of length in the model
equal to $0.5\times 10^3$km on the photosphere, good agreement between
the model results and observation is obtained up to the supergranule
cell size.  Deviations above that length scale may be due to several
causes: our assumption that the loops are perfectly semi-circular,
finite system size effects in the model or observations, or the force
free approximation used to calculate the flux tube connectivity from
observations of concentrations.  Comparing with
 the observed diffusive behavior of magnetic
concentrations\cite{diffusion_conc}  allows an additional
calibration of time. One unit of time in the model is 
equal to about 300 seconds on the photosphere.  From these three
calibrations we are able to determine values for the total solar flux
and the ``flux turnover time'', which both agree quantitatively with
observations. See Ref. [16] for details.

Our model predicts not only nominally universal
quantities like various
 critical exponents characterizing the flux network but
also quantities that have typical scales, such as total solar flux, the
distribution of footpoint separations, and the flux turnover time in the
corona. In order to represent the geometry of the coronal magnetic fields,
a three-dimensional model, as discussed here, is required.
Whether similar
 network models can be used to describe other high Reynolds number
astrophysical plasmas remains an open question.

\section{Seismicity}

Despite many efforts, seismicity remains an obscure phenomenon
shrouded in vague ideas without benchmarks of testability. At present, no
dynamical model can capture, simultaneously, the three
most robust statistical features of seismicity: (1) the
Gutenberg-Richter (GR) law\cite{gr_law,explain} for the distribution of earthquake magnitudes, and the clustering of activity in (2) space and (3)
 time. Spatio-temporal correlations include the
  Omori law\cite{omori94b,utsu95:_omori} for the decay in
the rate of aftershocks (see Eq.~4) and the fractal appearance of earthquake
epicenters. Note that stochastic processes like the ETAS process\cite{ETAS} require
 three or more power law distributions to be put in by hand. Since 
 these are the
 main scaling features     
  (1-3)   that we
 wish to establish a plausible dynamical mechanism for, ETAS models
 are not regarded by this author as  dynamical models of seismicity.
 To begin with, better methods to characterize seismicity
are needed.  Here I briefly discuss a network paradigm
put forward by Marco Baiesi and myself to this end.\cite{net_eq1,net_eq2}

\subsection{A Unified Approach to Different Patterns of Seismic Activity}

 Since seismic rates increase sharply after a large earthquake in the
region, events have been classified as aftershocks or main shocks, and
the statistics of aftershock sequences have been extensively studied.
Usually, aftershocks are collected by counting all events within a
predefined space-time window following a main 
event,\cite{gardner74:_after_window,keilis80:_after_window,knopoff00:_after_window}
These sequences are used e.g. to describe earthquake triggering\cite{helms02:_trig}
or predict earthquakes.\cite{kaganknopoff87:_alpha}
Obviously, some types of activity, such as
swarms, remote triggering,\cite{remote93} etc. cannot fit into this
framework. Perhaps a different description is needed for each pattern
of seismic activity.
On the other hand, it seems worthwhile to look for a unified
perspective to study various patterns of seismic activity
within a coherent framework.\cite{kagan94,bak02:_unified}

What if we do not fix {\it a priori}
the number of main shocks an event can be an aftershock of?  Perhaps
an event can be an aftershock of more than one predecessor. On the other hand,
all events are not equally correlated to each other. Probably the
situation is somewhere in between having one (or zero) correlated
predecessors or being strongly correlated to everything that happened
before.  In fact, a sparse but indefinite property of correlations between
events may be ubiquitous to all
intermittent spatio-temporal processes with memory.

 A sparse network (where each node is an event or earthquake) linking
strongly correlated pairs of events stands out as a good starting point for
describing seismicity in a unified way. In order to pursue this line
of reasoning, we 
 treat all
events on the same footing, irrespective of their magnitude, local
tectonic features, 
etc.\cite{bak02:_unified,corral03:_unified,davidsen_goltz04,davidsen_pac}  However, unlike other approaches we do not pre-define
any set of space or time windows. The sequence of activity itself selects
these.  Our method is also unrelated to that of Abe and Suzuki.\cite{abe}

\subsection{Relations Between Pairs of Events: The Metric}

We consider ONLY the relations between earthquakes and NOT the
 properties of individual events. Only catalogs that are considered
 complete are examined,\cite{davidsen_pac}
  and no preferred scales are imposed on the phenomenon.
Instead,  we invoke a
metric to estimate
the correlation between any two earthquakes, irrespective of how far apart
they are in space and/or time.\cite{net_eq1,net_eq2}
Consider as a null hypothesis\cite{jaynes_book} that earthquakes
are uncorrelated in time. Pairs of events where the null
hypothesis is strongly violated are correlated.
The metric measures the extent to which the null hypothesis is wrong.

The specific null hypothesis that we have investigated so
far\cite{net_eq1,net_eq2} is that 
earthquakes occur with a distribution of
magnitudes given by the GR law, with
epicenters located on a fractal of dimension $d_f$, randomly in time.
Setting $d_f=2$ does not change the observed scaling behaviors, nor does
varying the GR parameter, $b$.

An earthquake $j$ in the seismic region
occurs at time $T_j$ at location $R_j$.  Look backward in time to the
appearance of earthquake $i$ of magnitude $m_i$ at time $T_i$, at
location $R_i$.  How likely is event $i$, given that event
$j$ occurred where and when it did?  According to the null hypothesis, the number of
earthquakes of magnitude within an interval $\Delta m$ of $m_i$ that
would be expected to have occurred within the time interval
$t=T_j-T_i$ seconds, and within a distance $l=|R_i-R_j|$ meters, is
 
\beq 
n_{ij} \equiv (\textit{const}) \, t\, l^{d_f}  \, 10^{-b m_i}\, \Delta m \quad.
\label{eq:n}
\eeq 
Note that the space-time domain $(t,l)$ appearing in Eq.~\ref{eq:n} is self-selected by 
the particular history of  seismic activity in the region and not 
set by any observer. All earthquake pairs are considered on the same basis
according to this metric.

  Consider a pair of earthquakes $(i,j)$ where $n_{ij}\ll 1$; so that
the expected number of earthquakes according to the null hypothesis is
very small.  However, event $i$ actually occurred relative to $j$,
which, according to the metric, is surprising.  A small value
$n_{ij}\ll 1$ indicates that the correlation between $j$ and $i$ is
very strong, and {\it vice versa}. By this argument, the correlation
$c_{ij}$ between any two earthquakes $i$ and $j$ can be estimated to
be inversely proportional to $n_{ij}$, or \beq c_{ij}=1/n_{ij}
\quad.\eeq

We measured  $c_{ij}$ between all pairs of earthquakes greater
than magnitude 3 in the catalog
for Southern California from January 1, 1984 to December 31, 2003.\cite{catalog} The removal of small events assures that the catalog is complete, but otherwise
the cutoff magnitude is not important.
 The distribution of the correlation variables
$c_{ij}$ for all pairs ${i,j}$ was observed to be a power law
over fourteen orders of magnitude. Since no characteristic values of
$c$ appear in this distribution, it doesn't make sense to talk about
distinctly different
 classes of relationships between pairs. On the other hand,
due to the extremely broad distribution, each earthquake $j$ may have
exceptional events in its past with much stronger correlation to it than all
the others combined.  These strongly correlated pairs of events can be
marked as linked nodes, and the collection of linked nodes forms a
sparse network of disconnected, highly clustered graphs. 
 
\subsection{Directed, Weighted Networks of Correlated Earthquakes}

A sparse, directed, weighted network is constructed by only linking pairs
whose correlation $c$ exceeds a given threshold, $c_<$.  Each link is directed from
the past to the future.
For each threshold, $c_<$, the error made in deleting links with $c<c_<$ can be estimated. For instance,
throwing out 99.8\% of links gives results
accurate to within 1\%.  This leads to not only massive data reduction
with controllable error, but also a {\it renormalized model of seismicity},
which extracts the important, correlated degrees of freedom.

Each link contains several variables such as the
time between the linked events, the spatial distance between their
epicenters, the magnitudes of the earthquakes, and the
correlation between the linked pairs.  The networks are highly clustered with
a universal clustering coefficient $\approx 0.8$ for nodes with small degrees,
as well broad, approximately power law
 in- and out-degree distributions for the nodes.

Consequently, some events have many aftershocks, or outgoing links,
while others have one, or zero.  Also, some events are aftershocks
of many previous events, i.e. they have many incoming links,
 while others are aftershocks of only one (or zero) events.
The data reveal an absence
of characteristic values for the number of in or out-going links to an earthquake.\cite{net_eq2}
For each event $j$ that has at least one incoming link,  we define a link
weight to each "parent" earthquake
$i$ it is attached to as \beq w_{ij}= \frac{c_{ij}^{\eta}}{\sum_{k}^{\rm
in} c_{kj}^{\eta}} \quad , \label{eq:eta}\eeq where the sum is over
{all earthquakes $k$ with links going into $j$.  For instance, an event can be
$\frac{1}{2}$ an aftershock of one event, $\frac{1}{3}$
an aftershock of another, and $\frac{1}{6}$  an aftershock
of a third. Normally, the parameter
$\eta=1$, but it can also
 be varied without  changing the scaling properties
of the ensemble of network variables.

\subsection{The Omori Law for Earthquakes of All Magnitudes}

Fig.~\ref{fig:Omori} shows the rate of aftershocks for the Landers,
Hector Mine, and Northridge events.
The weights, $w$, of the links made  at time $t$
after one of these events are binned into geometrically increasing
time intervals.  The  total weight in each bin is then divided
by the temporal width of the bin to obtain a rate of weighted aftershocks
per second.  The same procedure is applied to each remaining event,
not aftershocks of these three.  An
 average is made for the rate of
aftershocks linked to events having a magnitude within an interval $\Delta m$
of $m$.  Fig.~\ref{fig:Omori} also shows the averaged results for $m=3$
(1871 events), $m=4$ (175 events), $m=5$ (28 events) and $m=5.9$ (4
events).

%FIG %%%
\begin{figure}[!tbp] 
\centering
\includegraphics[width=350pt,angle=0]{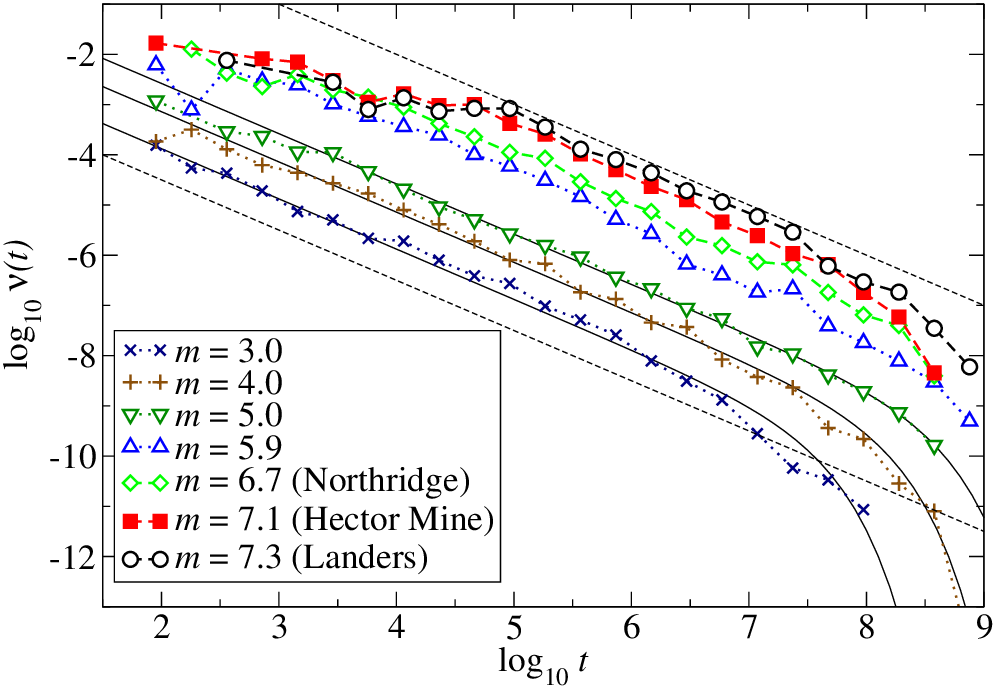}
\caption{The Omori law for aftershock rates. Rates are measured for 
aftershocks linked to earthquakes of different magnitudes.  
For each magnitude, the rate is consistent with the
Omori law, Eq.~\ref{eq:Omori2}. 
As guides to the eye, dashed lines represent a decay $\sim
1/t$. The dense curves represent the fits obtained by means of
Eq.~\ref{eq:fit} for $m=3$, $m=4$, and $m=5$.
\label{fig:Omori}}
\end{figure}
%FIG %%%
%FIG %%%

Earthquakes of all magnitudes have aftershocks that decay according
to an Omori law,\cite{omori94b,utsu95:_omori} \beq \nu(t) \sim
\frac{K}{c+t}\quad,\label{eq:Omori2} \quad {\rm for}\ t<t_{\rm
omori}\eeq where $c$ and $K$ are constant in time, but depend on the
magnitude of the earthquake.\cite{utsu95:_omori}  We find that the
 Omori law
persists up to time $t_{omori}$ that also depends on $m$.  The
function \beq \nu_m(t)\sim t^{-1}e^{-t/t_{\rm
omori}}\quad.\label{eq:fit}\eeq was fitted to the data, excluding
short times, where the the aftershock rates do not yet scale as $1/t$.
The short time deviation from power law behavior is presumably due to
saturation of the detection system, which is unable to reliably detect
events happening at a fast rate.  However, this problem does not occur
at later times, where the rates are lower.  Some examples of these
fits are also shown in Fig.~\ref{fig:Omori} for the intermediate
magnitude events.  From these fits, a scaling law \beq
t_{\rm omori}(m) \simeq 10^{5.25+0.74 m}\quad {\rm sec} \label{eq:t_cutoff} \eeq was observed for times shorter than the
duration of the catalog.  It  corresponds to $t_{\rm
omori}\approx 11$ months for $m=3$, and to $t_{\rm omori} \approx 5$
years for $m=4$.  An  extrapolation yields $t_{\rm omori} \approx 1400$
years for an event with $m=7.3$ such as the Landers event and $t_{\rm
omori} \approx 26,000$ years for the 9.0 Northern Sumatra earthquake
causing the 2004 Asian tsunami.  These results
 confirm Kagan's conjecture that
aftershocks can rumble on for centuries.\cite{kagan_quote} Indeed, with previous
measurement techniques it was not possible to test his
hypothesis.

\section{Acknowledgments}
The author thanks David Hughes, Marco Baiesi, and J\"orn Davidsen
 for enthusiastic
 discussions and their collaborative efforts which
contributed to the work discussed here, as well as Peter Grassberger for
critical comments on the manuscript.  She also thanks her colleagues at
the Perimeter Institute, including
Fotini Markopoulou and Lee Smolin, for wide ranging
conversations.

\end{document}